\documentclass[11pt,onecolumn,amssymb,nofootinbib]{revtex4}
\usepackage{amsmath, amsthm, amscd, amssymb}
\usepackage{graphicx, braket}
\usepackage{bm}
\usepackage{bbm}
\usepackage{epstopdf}

\begin{document}

\title{\bf Are Astrophysical ``Black'' Holes Leaky?} \bigskip

\author{Stephen L. Adler}
\email{adler@ias.edu} \affiliation{Institute for Advanced Study,
1 Einstein Drive, Princeton, NJ 08540, USA.}

\begin{abstract}
 We continue a  study by Adler and Ramazano\u glu (AR) of ``black'' holes as modified by a scale invariant dark energy action.  For the spherically symmetric Schwarzschild-like case, (AR) found that there is no event horizon; hence spacetime is not divided by the ``black'' hole into causally disconnected regions.  We review the formalism for locating trapped surfaces and apparent horizons, and show that the modified ``black'' hole has no trapped surfaces.  Thus one suspects that it is ``leaky'', and that there will be a ``black hole wind'' of particles streaming out from the location of the nominal horizon.  This will have astrophysical consequences, for example, the wind may feed and stabilize star formation in the vicinity of the ``black'' hole.     We initiate a study of the stationary axially-symmetric rotating ``black'' hole as modified by a scale invariant dark energy action, i.e, one that is Kerr-like.   To set up the axial case, we note that the conserving completion of the dark energy stress energy tensor can be calculated algebraically by solving a $2 \times 2$ matrix equation. Using Mathematica we calculate and simplify the modified system of Einstein equations for the axial case in quasi-isotropic coordinates, and give appropriate boundary conditions.   Solution of these equations, which may require developing a special purpose numerical program,  is rendered tricky by a residual general coordinate invariance of quasi-isotropic coordinates.  This leads to the interesting mathematical question of representing a positive planar function as the gradient squared of a harmonic function.
\end{abstract}

\maketitle
\section{Introduction}

\subsection{Mathematical black holes versus astrophysical ``black'' holes}

Mathematical black holes, as described in the magisterial review of Chandrasekhar \cite{chandra}, are solutions of the Einstein field equations characterized by just two parameters, the mass $M$ and the angular momentum per unit mass $a$.  They have no other internal parameters, and possess an event horizon that divides spacetime into two causally disconnected regions, the interior, which is inaccessible to observers and from which nothing can escape, and the exterior, which is observable.

There are now many astrophysical observations implying the presence in the universe of extremely compact objects, which are interpreted as black holes.  But are they true mathematical black holes with an event horizon, or modified forms of the idealized black hole, that may lack an event horizon and therefore have observable interiors?  This question has been raised by a number of authors, and is at the root of the present paper, which focuses on a compact object model based on a novel form of dark energy, as described in the following subsection.  Other proposals for modified holes are discussed in a recent review by Cardoso and Pani \cite{pani}, who give the name ``exotic compact objects'' to a class of such models, and focus on possible signatures in gravitational radiation for the lack of a horizon .  Our focus is somewhat different, on the possibility of particle leakage from the compact object, which may have implications for astrophysical phenomena such as star formation and the origin of jets.

But before proceeding, first an observation about notation.  If particles can leak out of a modified hole, it is {\it not} black, and so referring to it as a black hole is a bit of a non sequitur.  To clarify this semantic issue  we use in this article the term ``black'' hole, with the quotes denoting an object which may have {\it no} event horizon, but which otherwise appears to astronomers very similar to the idealized mathematical black hole.  For mathematical black holes we continue to omit the quotes.

\subsection{Motivations for a  Weyl scaling invariant dark energy action}

In papers over the last eight years \cite{adler1}-\cite{adler6} we have explored the postulate that the part of the gravitational action that depends only on the undifferentated metric $g_{\mu\nu}$, but involves no metric derivatives, is invariant under the Weyl scaling $g_{\mu\nu}\to \lambda g_{\mu\nu}$. Adoption of this postulate implies that the so-called  ``dark energy'' action has the  three-space general coordinate invariant,  but frame-dependent,  form
\begin{equation}\label{eff}
S_{\rm eff}=-\frac{\Lambda}{8 \pi G} \int d^4x ({}^{(4)}g)^{1/2}(g_{00})^{-2}~~~,
\end{equation}
rather than the usually assumed vacuum energy form
\begin{equation}\label{vacen}
S_{\rm eff}=-\frac{\Lambda}{8 \pi G} \int d^4x ({}^{(4)}g)^{1/2}~~~,
\end{equation}
where $\Lambda$ is the observed cosmological constant, and ${}^{(4)}g=-\det(g_{\mu\nu})$.
Since the unperturbed Friedmann-Lema\^itre-Robertson-Walker (FLRW) cosmological metric has  $g_{00}=1$, in this context the action of Eq. \eqref{eff} mimics the standard cosmological constant action of  Eq.  \eqref{vacen}, but when $g_{00}$ deviates from unity, their consequences differ. There are a number of motivations for studying the possibility that dark energy arises from the action of Eq. \eqref{eff}.  First, we were led to it by a study \cite{adler1} of the induced gravitational action in our proposed pre-quantum theory based on
trace dynamics, where the properties of Weyl scaling invariance and frame dependence of the underlying theory are inherited by the induced
action.  Second, quite independently 't Hooft \cite{thooft} has advocated that Weyl scaling invariance is ``the missing  symmetry component of space and time'', and a natural place to test this suggestion is in the dark energy action,  the form of which is still a subject of investigation, unlike the well-tested Einstein-Hilbert gravitational action and the standard model particle action.  Finally, assuming that dark energy arises as a vacuum energy from the action of Eq. \eqref{vacen} leads to the cosmological constant fine tuning problem, which is not implied by alternative forms of the dark energy action.  We have recently completed a mini-review \cite{adreview} of our papers \cite{adler1}-\cite{adler6}, which includes an extended discussion of motivations for our postulate, to which the reader is referred for further details.

Our more recent papers \cite{adler5}, \cite{adler6} have been devoted to exploring cosmological implications of Eq. \eqref{eff}  at the level of first order cosmological perturbations from a
FLRW background. The aim of the current paper is to return to a study of the implications of the action of Eq. \eqref{eff} for the physics of ``black'' holes, begun in the paper \cite{adler2} co-authored with Ramazano\u glu, referred to henceforth as (AR).  An extension of (AR) forms the subject of Section II of this paper, and our principal new result is that for a Schwarzschild ``black'' hole as modified by the dark energy action of Eq. \eqref{eff} \big(termed ``Schwarzschild-like'' in (AR)\big), there is not only no event horizon, but also no apparent horizon and no trapped surfaces. Based on
this, we conjecture (but do not prove) that there will be a ``black hole wind'' emanating from the hole, and briefly give some possible astrophysical applications.  In Section III we turn to extending our spherical results to the axially symmetric case, corresponding to a Kerr ``black'' hole as modified by the dark energy action of Eq. \eqref{eff} (which we term ``Kerr-like'').  Here our principal new result is to set up the Einstein equations for the metric components as modified by Eq.\eqref{eff}, in an explicit form suitable for numerical analysis.  We analyze a residual general coordinate invariance of axial quasi-isotropic coordinates and its mathematical implications, and related to this, we discuss  difficulties encountered in an attempted numerical solution by the Mathematica differential equation solver.   A numerical solution will be needed to see if there are results for the horizon structure analogous to those found in the spherical case, but it appears that a special purpose program may have to be developed in the future to complete the numerical work in the axial case.  In Section IV we give a short list of problems for future investigation, which is elaborated on in the review \cite{adreview}.  In the Appendix, we give
some formulas that were useful in checking the axial vector equations.

\section{Schwarzschild-like ``black'' holes}
In (AR) we gave analytic and numerical results of solving the Einstein equations arising from the Einstein-Hilbert action coupled to the action of Eq. \eqref{eff} in the spherically symmetric case, giving rise to what we termed ``Schwarzschild-like'' ``black'' holes.    We found that, as anticipated from the factor of $g_{00}^{-2}$ in Eq. \eqref{eff}, the horizon structure is substantially changed. At distances much larger than $2 \times 10^{-18} {\cal M}^2$ cm from the nominal horizon, with ${\cal M}$ the ``black'' hole mass in solar mass units, the solutions closely resemble the standard Schwarzschild solution until cosmological distances are reached. Within $2 \times 10^{-18} {\cal M}^2$ cm from the nominal horizon, the behavior of $g_{00}$ changes, with $g_{00}$ remaining nonvanishing until an internal singularity is reached.  In spherical coordinates, $g_{00}$ has a square root branch point in the vicinity of the nominal horizon, which is a coordinate singularity.  In isotropic coordinates, $g_{00}$ remains nonzero and real analytic from cosmological distances down to the internal physical singularity.  There is no event horizon in the modified ``black'' hole, and no division of physical space into causally disconnected regions.  However, in (AR) we did not
examine whether there is an ``apparent'' horizon in the modified ``black'' hole, defined as the outer boundary of trapped surfaces, or in technical terms, the outermost marginally trapped surface \cite{hawking}.  The presence or absence of trapped surfaces is what determines whether material can leak out of the ``black'' hole. Studying the apparent horizon structure of spherical,
Schwarzschild-like ``black'' holes, using the numerical solution in isotropic coordinates found in (AR),  is the topic to which we turn next.\footnote{There is no closed form analytic solution in the Schwarzschild-like case.  In addition to a numerical solution, there are only approximate analytic results such as a few orders of  expansions
in powers of $\Lambda$, as developed in detail in (AR).  This contrasts with the case when the dark energy action is given by Eq. \eqref{vacen}, which yields the closed form analytic Schwarzschild-de Sitter metric, and has a horizon surrounding the hole with similar properties to the Schwarzschild case.  }

\subsection{Theory of trapped surfaces and apparent horizons}

We now summarize the theory of locating  trapped surfaces and apparent horizons, as discussed in the reviews \cite{faraoni} and \cite{krishnan1}, following closely the extension given in our article \cite{adler7}.  This is relevant to the issue of ``black'' hole ``leakiness''  because when there are no trapped surfaces, and no event or apparent horizons,  it is no longer forbidden for material to flow  outwards from the modified ``black'' hole.

For a  compact, orientable  2-surface embedded in 4-space, there are two orthogonal directions corresponding to outgoing and ingoing null rays, with respective tangents $\ell^{\nu}$ and $n^{\nu}$ respectively.  Let $g_{\mu\nu}$ be the metric \big(for which we take the $(-,+,+,+)$ convention\big) and $g^{\mu\nu}$ its inverse. We begin by forming the projector onto the 2-surface,
\begin{equation}\label{proj1}
h^{\mu\nu}=g^{\mu\nu}+\frac{\ell^{\mu}n^{\nu}+\ell^{\nu}n^{\mu}}{-\ell^{\alpha}n_{\alpha}}~~~,
\end{equation}
which if we adopt \cite{nielsen} the convenient normalization convention
\begin{equation}\label{normconv}
\ell^{\alpha}n_{\alpha}=-2~~~,
\end{equation}
becomes
\begin{equation}\label{proj2}
h^{\mu\nu}=g^{\mu\nu}+\frac{\ell^{\mu}n^{\nu}+\ell^{\nu}n^{\mu}}{2}~~~.
\end{equation}
By construction,  $h^{\mu\nu}$ projects the null vector normals to zero,
\begin{equation}\label{zero}
h^{\mu\nu}\ell_\mu =h^{\mu\nu}\ell_\nu =h^{\mu\nu}n_\mu =h^{\mu\nu}n_\nu = 0~~~.
\end{equation}
Evidently  the projector $h^{\mu\nu}$ is invariant under reciprocal rescalings of $\ell^{\nu}$ and $n^{\nu}$ according to
\begin{equation}\label{rescaling}
\ell^\nu \to \kappa \ell^\nu~,~~~n^\nu \to \kappa^{-1} n^\nu~~~.
\end{equation}
A rescaling of $\ell^\nu$ and $n^{\nu}$ with constant $\kappa$ has been considered previously by Ashtekar, Beetle, and Lewandowski \cite{ashtekar},  Ashtekar and Krishnan \cite{krishnan2} , and Krishnan \cite{krishnan1}, but here we allow $\kappa=\kappa(x)$ to be a general
nonconstant scalar function of the spacetime coordinate $x$.

The {\it expansion}  $\theta_{\ell}$ of a bundle (or congruence) of null rays associated with the tangent vector $\ell$  is a measure of the fractional change of the cross sectional area of the bundle as one moves along the central ray of the bundle.  Using the projector $h^{\mu\nu}$, the expansions $\theta_\ell$ and $\theta_n$ associated with the outgoing and ingoing null vectors $\ell^{\nu}$ and $n^{\nu}$ are calculated from the formula
\begin{equation}\label{expansion}
\theta_\ell=h^{\mu\nu} \nabla_\mu\ell_\nu~~,~~~~
\theta_n=h^{\mu\nu} \nabla_\mu  n_\nu~~~,
\end{equation}
with $\nabla_\mu$ as usual the covariant derivative.
To see how these expansions transform under the rescalings of Eq. \eqref{rescaling}, we note that
\begin{equation}\label{nabla1}
\nabla_\mu \kappa(x) \ell_\nu = \ell_\nu \partial_\mu \kappa(x) + \kappa(x) \nabla_\mu \ell_\nu~~~,
\end{equation}
with $\partial_\mu$ the partial derivative.
Since the inhomogeneous term $\ell_\nu \partial_\mu \kappa $ is projected to zero by $h^{\mu\nu}$ by virtue of Eq. \eqref{zero}, the expansion
$\theta_{\ell}$ transforms under Eq. \eqref{rescaling} by the simple scaling formula
\begin{equation}\label{thetatrans1}
\theta_\ell \to \kappa \theta_\ell~~~,
\end{equation}
and similarly, $\theta_n$ transforms under the reciprocal scaling formula
\begin{equation}\label{thetatrans2}
\theta_n \to \kappa^{-1} \theta_n~~~,
\end{equation}
with the product $\theta_\ell \theta_n$ invariant
\begin{equation}\label{prodtrans}
\theta_\ell \theta_n \to \theta_\ell \theta_n~~~.
\end{equation}
Thus calculations of the expansions $\theta_\ell$, $\theta_n$ using the standard recipe have a covariance group under the transformations of Eq. \eqref{rescaling} with general non-constant $\kappa(x)$, with the associated product of Eq. \eqref{prodtrans} an invariant.

Consider now what happens if we compute the expansions $\theta_{\ell,n}$ for the same physics viewed from different choices of coordinates.
In each coordinate system we have to pick null vectors $\ell$ and $n$, and different ways of doing this that satisfy the norm convention of
Eq. \eqref{normconv} will differ by the rescaling freedom of Eq. \eqref{rescaling}.  Thus, if we pick the most convenient definitions of $\ell$ and $n$ in each coordinate system, for example those with equal time components and opposite signs of the spatial components, we will in general get {\it different} values of the expansions $\theta_\ell$, $\theta_n$ in the various coordinate systems, with only the product $\theta_\ell \theta_n$ the same in all systems.  However, the value of this product is in itself a useful diagnostic.  According to the usual classification reviewed in \cite{faraoni}--\cite{nielsen}, $\theta_\ell \theta_n<0$ corresponds to a normal or untrapped surface; $\theta_\ell \theta_n>0$ corresponds to a trapped or antitrapped surface; and $\theta_\ell\theta_n=0$ corresponds to a marginal surface, such as the case $\theta_\ell=0$, $\theta_n>0$ which defines the future apparent horizon.  Thus an apparent horizon can be located by computing $\theta_\ell \theta_n$ in any coordinate system, even though the individual values of $\theta_\ell$ and $\theta_n$ may vary.

As an illustration of this formalism, let us calculate $\theta_\ell$ and $\theta_n$ in spherical coordinates.  We write the general static spherically symmetric line element in  the form,
\begin{equation}\label{Selt}
 ds^2=-F(r)^2 dt^2 + G(r)^2 dr^2 + r^2 d\theta^2 + r^2 \sin^2\theta d\phi^2~~~,
 \end{equation}
 corresponding to $g_{tt}=-F(r)^2$ and $g_{rr}=G(r)^2$.
 Making the convenient and symmetrical choice
 \begin{align}\label{nv2}
\ell^\nu=&\big(1/F(r),1/G(r),0,0\big) ~~~,\cr
n^\nu=&\big(1/F(r),-1/G(r),0,0\big)  ~~~,\cr
\end{align}
which obey $\ell^\alpha  \ell_\alpha = n^\alpha  n_\alpha =0$ and $\ell^\alpha   n_\alpha =-2$, a Mathematica calculation  using the recipe of Eq. \eqref{expansion} gives
 \begin{align}\label{Sexpan}
 \theta_{\ell}=& \frac{2}{rG(r)}~~~,\cr
 \theta_{n}=& -\theta_\ell~~~,\cr
 \theta_\ell \theta_n =& -\frac{4}{r^2G(r)^2}~~~.\cr
 \end{align}
For a Schwarzschild black hole, $G(r)^2=1/F(r)^2=(1-2M/r)^{-1}$, so Eq. \eqref{Sexpan} becomes
\begin{equation}\label{schwarz}
\theta_\ell \theta_n = -\frac{4}{r^2}(1-2M/r)~~~.
\end{equation}
This is negative for $r>2M$, zero at $r=2M$, and switches to positive for $r<2M$.  Thus the spherical surfaces outside $2M$ are untrapped, and those within $2M$ are trapped, with an apparent horizon (the outer boundary of trapped surfaces) at $2M$.  So in this case we get the expected result that the apparent horizon coincides with the event horizon at $r=2M$.

\subsection{Absence of an apparent horizon in the Schwarzschild-like case}

We now apply this recipe to see whether there is
an apparent horizon, and whether there are any trapped surfaces, in the Schwarzschild-like ``black'' hole studied in (AR).  We start now from the
 general static spherically symmetric line element written in isotropic coordinates,
\begin{equation}\label{lineltiso}
ds^2=-\frac{B(r)^2}{A(r)^2} dt^2 +\frac{A(r)^4}{r^4}\big(dr^2 +r^2 d\theta^2 + r^2 \sin^2\theta d\phi^2\big)~~~.
\end{equation}
To slightly simplify the calculation of the expansions, we define $F(r)=B(r)/A(r)$, $H(r)=A(r)^2/r^2$, in terms of which the line element becomes
\begin{equation}\label{lineltiso1}
ds^2=-F(r)^2 dt^2 +H(r)^2\big(dr^2 +r^2 d\theta^2 + r^2 \sin^2\theta d\phi^2\big) ~~~.
\end{equation}
Suitable null vectors $\ell^\nu$ and $n^\nu$ satisfying the normalization convention of Eq. \eqref{normconv} are
\begin{align}\label{null}
\ell^\nu=&\big(1/F(r),1/H(r),0,0\big) ~~~,\cr
n^\nu=&\big(1/F(r),-1/H(r),0,0\big)  ~~~,\cr
\end{align}
and substituting these into the recipe of Eq. \eqref{expansion} gives
\begin{align}\label{expiso}
\theta_\ell= &2 \frac{H(r)+rH'(r)}{rH(r)^2}= \frac{2r}{A(r)^3}[2rA'(r)-A(r)]~~~,\cr
\theta_n=&-\theta_{\ell}~~~.\cr
\end{align}
Hence
\begin{equation}\label{expiso1}
\theta_\ell\theta_n  = -\theta_\ell^2=-\frac{4r^2}{A(r)^6}[2rA'(r)-A(r)]^2 \leq 0.
\end{equation}

Let us first apply Eq. \eqref{expiso1} to the case of an ordinary Schwarzschild black hole viewed in isotropic coordinates, for which
\begin{equation}\label{schwarz1}
B=r-M/2\,,~A=r+M/2\,,~A'=1\,,~2rA'-A=r-M/2~~~.
\end{equation}
In these coordinates the event horizon is at $r=M/2$, where $B$ vanishes.  Since the mapping relating the spherical radial coordinate $\bar r$,  \big(which we denoted as $r$ in Eqs. \eqref{Selt}-\eqref{schwarz} of the previous section\big) to the isotropic radial coordinate $r$ of this section is
\begin{equation}\label{map}
\bar r= r\left(1+\frac{M}{2r}\right)^2=r+M+\frac{M^2}{4r}~~~,
\end{equation}
at $r=M/2$ one has $\bar r=2M$, giving the spherical coordinate event horizon location.  However, we see that the isotropic coordinate domain $0<r<\infty$ gives a double covering of the spherical coordinate domain $2M<r<\infty$, and never reaches the interior of the Schwarzschild black hole as
described in spherical coordinates.  That is why $\theta_\ell \theta_n$ of Eq. \eqref{expiso1} never attains positive values, and so only describes the untrapped surfaces in the exterior of the Schwarzschild black hole.

We have gone through this detail in the Schwarzschild case, to emphasize that when using isotropic coordinates in the Schwarzschild-like case of a spherical ``black'' hole as modified by the scale invariant dark energy action of Eq. \eqref{eff}, we must establish whether these
coordinates give a covering of the interior region as well as the exterior region, or only a double covering of the exterior region.  In Section 3 of (AR) a numerical analysis of Schwarzschild-like ``black'' holes was given, after introducing dimensionless variables by rescalings  $\Lambda \to 1$, $r \to x=\Lambda^{1/2}r$, and $M \to \hat M = \Lambda^{1/2} M$.  For details of the Einstein equations and boundary conditions that were integrated using these rescalings see (AR).  Sample results for a rescaled mass $\hat M=10^{-6}$ were obtained by using the Mathematica integrator NDSolve, which indicated the presence of singularities at $x_{\rm lower}=5.000061\times 10^{-7}$ and $x_{\rm upper} =0.8276970$.  We interpreted the singularity at $x_{\rm upper}$ as the singularity at $\infty$ found in (AR) in polar coordinates, i.e.,  a singularity at cosmological distances, and we interpreted the singularity at $x_{\rm  lower}$ as the singularity interior to  the ``black'' hole.  This interpretation is supported by computing the proper distance between the singularities,
\begin{equation}\label{proper}
D=\Lambda^{-1/2} \int_{x_{\rm lower}}^{x_{\rm upper}} dx \frac{A(x)^2}{x^2}=0.927374 \Lambda^{-1/2}~~~,
\end{equation}
in excellent agreement with the value $D=0.927371 \Lambda^{-1/2}$ for the proper distance from the polar radius $\bar r=0$ to $\bar r=\infty$
computed using the zero mass limit of the ``master equation'' derived in the spherical case.  Thus the isotropic coordinate calculation  gives a single covering of the region extending from the interior singularity of the Schwarzschild-like ``black'' hole to to the singularity at
cosmological distances, and not a double covering of the exterior region of the ``black'' hole.

We can now turn to discuss whether the Schwarzschild-like ``black'' hole has horizons.  In Fig. 5 of (AR) we exhibited a plot of $g_{00}(x)$ in
isotropic coordinates, showing that there is no event horizon between $x_{\rm lower}$ and $x_{\rm upper}$.  From Eq. \eqref{expiso1}, we
see that the product $\theta_\ell \theta_n \leq 0$, and so there are no trapped surfaces where this product is positive.  A plot of
$\theta_\ell$ versus $x-x_{\rm lower}$ is given here in Fig. 1, which shows that $\theta_{\ell}$ vanishes only at $x-x_{\rm lower}\simeq 1.4097 \times
10^{-9}$.  But this cannot be termed an apparent horizon because the spherical surfaces lying within it are untrapped, so it is not an outer boundary of trapped surfaces.  The proper distance from $x_{\rm lower}$ to $x_{\rm lower}+1.4097 \times 10^{-9}$ is $5.6347 \times 10^{-9} \Lambda^{-1/2}$, a minute fraction of the proper distance $D=0.927374 \Lambda^{-1/2}$ between $x_{\rm lower}$ and $x_{\rm upper}$.\footnote{Since the zero in $\theta_{\ell}$ is so close in proper distance to the singularity at $x_{\rm lower}$, and is absent for some choices of formally equivalent differential equation systems, it may well be a computational artifact.}  So in the analysis of $\theta_\ell \theta_n$ there is again no indication of the double covering found in the unmodified Schwarzschild case.
Thus we conclude that unlike the Schwarzschild black hole, which has an apparent horizon coinciding with its event horizon,
 the Schwarzschild-like ``black'' hole has no event horizon, no apparent horizon, and no trapped surfaces.

 \begin{figure}[t]
\begin{centering}
\includegraphics[natwidth=\textwidth,natheight=300,scale=1.8]{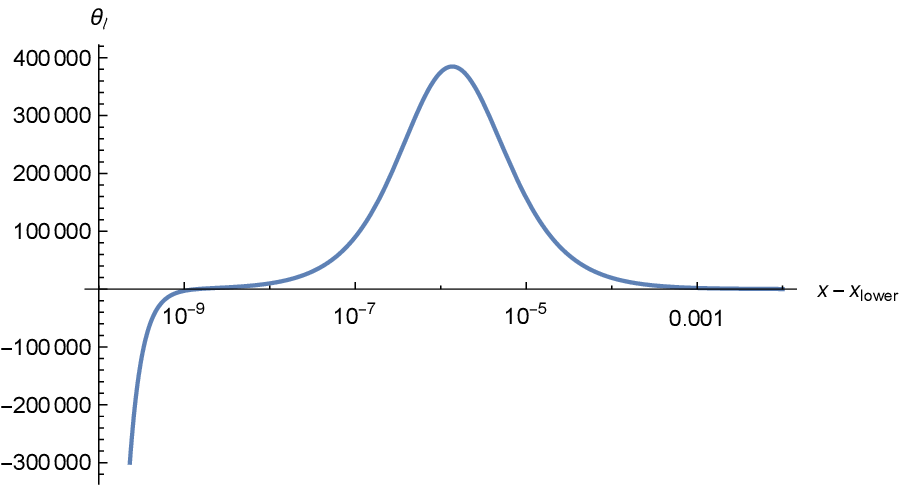}
\caption{$\theta_\ell$ versus $x-x_{\rm lower}$ }
\end{centering}
\end{figure}

\subsection{A ``black hole wind'' and possible astrophysical consequences}

For a Schwarzschild-like ``black'' hole with no interior trapped surfaces, the possibility is thus opened for there to be a leakage of particles from the interior to outside the nominal horizon.  This  flux would be quite distinct from the quantum mechanical Hawking radiation flux, and could be orders of magnitude larger.  Therefore we postulate that emerging from ``black'' holes, as modified by the dark energy action of Eq. \eqref{eff}, there will be a ``black hole wind'' of particles with enough velocity to escape to an astronomical distance from the hole.  An important issue for the future will be extending the calculation of the preceding section to include normal matter that has entered the ``black'' hole, but is not permanently trapped, and developing the physical concepts needed to calculate the flux of particles emerging from the hole. So we do not claim to have ``proved'' the existence of a ``black'' hole wind, only to have established that it is plausible possibility given the modifications in horizon structure induced by the action of Eq. \eqref{eff}.
If there is a  ``black'' hole wind, it could have interesting astrophysical consequences.  We here note two of them;  see also the discussion in the review \cite{adreview}.

\begin{itemize}

\item  Observations of the central  ``black'' hole in our galaxy show the presence of young stars in its close vicinity.  Lu et al. open their article \cite{lu} by stating that: ``One of the most perplexing problems associated with the supermassive black hole at the center of our Galaxy is the origin of young stars in its close vicinity''.  Similar clusters of young stars are found in the vicinity of  supermassive ``black'' holes in nearby galaxies \cite{lu1}.   A ``black'' hole wind of particles could possibly furnish a mechanism for
    the origin of such stars, both by providing material for their formation, and by providing an outward pressure cushioning nascent stars  against the gravitational tidal forces of the ``black'' hole.  This idea could be tested in a phenomenological way, by postulating parameters for the ``black'' hole wind (particle type, velocity, flux) and seeing if they can account for young star formation \cite{ak}.  Related considerations apply to the possibility of galaxy formation catalyzed by a central leaky ``black'' hole \cite{gal}.

\item The presence of highly collimated jets that emerge from active galactic nuclei, which are believed to contain supermassive ``black'' holes, is a striking feature that has attracted much attention.  The prevailing models for the formation of these jets, in particular the Blandford-Znajek \cite{bland} mechanism, assume no material (other than the negligible Hawking radiation flux) emerges from the horizon of the central hole, as expected for an idealized mathematical black hole.    If this assumption is incorrect, and there is a ``black'' hole wind at the classical physics level, this wind could play a contributing role in the formation of the observed jets.  Thus it will be interesting to solve the rotating ``black'' hole analog of the calculations of the preceding section, to get the structure of a rapidly rotating Kerr-like ``black'' hole, to see if there is evidence for preferential emission of a ``black'' hole wind along the directions of the rotation axis.

    \end{itemize}

\section{Kerr-Like ``Black'' Holes}

In Section II we discussed the horizon structure of modified spherically symmetric ``black'' holes.  But just as  Schwarzschild black holes are a special case of axially-symmetric, rotating Kerr black holes, we expect there to be modified axially symmetric ``Kerr-like'' ``black'' holes  generalizing the spherically symmetric ``Schwarzschild-like'' ``black'' holes studied in (AR) and in Section II.  To get analogous results for these, we will first have to generalize to the axial case the analytic and numerical analysis of (AR), the topic to which we now turn.

\subsection{Setting up the axial case:  Quasi-isotropic coordinates, conserving completion of the stress-energy tensor, Einstein equations, and  boundary conditions}

A sufficiently general line element for a stationary axially symmetric metric  \cite{chandra}  is
 \begin{align}\label{statmetric}
 ds^2=&-e^{2\nu} dt^2 + e^{2\psi}(d\phi-\omega dt)^2 + e^{2\mu} (d\rho^2 + dz^2)~~~\cr
 =& (- e^{2\nu}+\omega^2 e^{2\psi}) dt^2 - 2 e^{2\psi} \omega dt d\phi + e^{2\psi} d\phi^2 + e^{2\mu} (d\rho^2 + dz^2)~~~,\cr
 \end{align}
 with the four quantities $\nu,\psi,\mu,\omega$ functions of $\rho$ and $z$. We have written this line element in terms of cylindrical
 coordinates $\rho$ and $z$, which are related to the spherical coordinates $r$ and $\theta$ used in Eq. \eqref{lineltiso} by
 \begin{equation}\label{transfromspher}
 \rho=r\sin\theta ~,~~~z=r \cos\theta~~~,
 \end{equation}
 from which we see that
 \begin{equation}\label{rhozline}
 d\rho^2 + dz^2=dr^2 + r^2 d\theta^2~~~.
 \end{equation}
 Thus, as in Eq. \eqref{lineltiso}, the line element differentials in the $\rho-z$ or equivalently $r-\theta$ plane have a common coefficient, which is why
Eq. \eqref{statmetric} is the axial analog of the  spherical isometric line element of Eq. \eqref{lineltiso}, and why it is termed quasi-isometric.  The Kerr metric can be readily transformed \cite{brandt}, \cite{allen} from spherical Boyer-Lindquist coordinates to the quasi-isometric coordinates of Eq. \eqref{statmetric}.

 Setting up the Einstein equations arising from the Einstein-Hilbert action plus the dark energy action of Eq. \eqref{eff} proceeds in a number of steps, all of which we carried out using the algebraic calculation tools of Mathematica.  We summarize the steps in outline form, and then state the final results.

 \begin{itemize}
 \item  As discussed in detail in \cite{adler1} and in \cite{adreview}, since the dark energy action of Eq. \eqref{eff} is only three-space general coordinate invariant, its variation with respect to the full metric $g_{\mu\nu}$ does not yield a covariantly conserved stress-energy tensor to serve as the source term for the Einstein equations.\footnote{An action gives consistent equations of motion when varied with respect to independent dynamical variables.  The 10 components of the full metric $g_{\mu\nu}$ are not independent, because they should be restricted by coordinate conditions.  However, when the action is four-space general coordinate invariant,  one gets the correct energy-momentum tensor by varying with respect to the full $g_{\mu\nu}$, treating all 10 components as independent.  This works  because the action  is invariant under general coordinate transformations, and so changes in the coordinate conditions reducing the 10 components to 6 independent ones produce an action variation of zero.}  The correct way to
     use Eq. \eqref{eff} is to vary it with respect to the  six independent spatial
     metric components $g_{ij}$, giving the spatial components of the stress-energy tensor $T^{ij}$.  When the Einstein-Hilbert action is included, varying with respect to $g_{ij}$ gives the spatial components of the Einstein equations $G_{ij}=-8\pi G T_{ij}$.  Then one can determine the remaining components $T^{0j}=T^{j0}$ and $T^{00}$ by solving the  differential equations for covariant conservation (as dictated by the Bianchi identities obeyed by $G_{\mu\nu}$, which require  $\nabla_{\mu}T^{\mu\nu}=0$), a procedure that we term covariant completion. Covariant derivatives are conveniently calculated using a special purpose Mathematica notebook for general relativity.\footnote{We obtained this ``folklore'' notebook from Fethi  Ramazano\u glu in the course of the work on \cite{adler2}, and extended it to check Bianchi identities and covariant conservation, in addition to its original purpose of computing the Einstein tensor.} In the
     spherically symmetric ``black'' hole case studied in \cite{adler2}, covariant completion reduces to an algebraic equation in one variable $T^{tt}\equiv T^{00}$, which can be immediately solved.  In the axially-symmetric case with the line element of Eq. \eqref{statmetric}, covariant completion reduces to a set of simultaneous algebraic equations in the two unknowns $T^{tt}$ and $T^{t \phi}=T^{\phi t}$, which are readily solved using the coefficient extraction and matrix inversion functions of Mathematica.

   \item Using the Mathematica notebook for general relativity, the Einstein tensor corresponding to the metric of Eq. \eqref{statmetric} is readily generated.  Only four components of the Einstein tensor are needed to give four equations to determine the four metric functions $\nu,\psi,\mu,\omega$, and we take these as $G_{tt}$,\, $G_{\rho\rho}+G_{zz}$,\,$G_{\phi\phi}$,\, and
       $G_{t\phi}$.  The Einstein equations then follow from equating $G_{\mu \nu}$ to the full stress-energy tensor term $-8 \pi G T_{\mu\nu}$ obtained by covariant completion.
   \item  We find that the resulting equations take a slightly simpler form when linear combinations are taken which isolate in separate  equations the second derivatives $\nabla^2 \nu$, $\nabla^2 \psi$, $\nabla^2 \mu$, and $\nabla^2 \omega$, plus terms with first derivatives or no derivatives of the unknown functions, with $\nabla^2$ the two-dimensional Laplacian
       \begin{equation}\label{laplacian}
       \nabla^2 \equiv (\partial_\rho)^2 + (\partial_z)^2~~~.
       \end{equation}
       This is accomplished by further use of the matrix operations of Mathematica.

 \end{itemize}

 The result of these Mathematica calculations is a set of four second-order differential equations to be solved in the
 axially symmetric case.  In these equations, the terms with no coefficient $\Lambda$ come from varying the Einstein-Hilbert action by itself,  and are obeyed  by the Kerr metric rewritten in quasi-isotropic coordinates given in Appendix A.1.  The terms with   $\Lambda$ as coefficient arise from varying the Weyl scaling invariant dark energy action of Eq. \eqref{eff} with respect to the spatial metric components, and then carrying out a conserving completion to give the full stress energy tensor that is covariantly conserved, as required for consistency by the Bianchi identities.  Solving the following set of equations numerically will give the axial rotating analog of the spherical calculation performed in (AR), and will permit study of horizon structure (the event horizon, apparent horizon, and presence or absence of trapped surfaces) in the Kerr-like case:

\begin{align}\label{Omeg}
\nabla^2 \omega=&\partial_\rho \omega (\partial_\rho \nu-3 \partial_\rho \psi)+
\partial_z\omega (\partial_z \nu-3 \partial_z \psi)\cr
&+\Lambda\big[-8 \exp(2\mu + 2\nu)(\omega/D_1)[1+\omega (\partial_\rho \psi \partial_z \nu - \partial_\rho \nu \partial_z \psi)/D_2]\big] ~~~ ,\cr
\end{align}

\begin{align}\label{mu}
\nabla^2 \mu=&\partial_\rho\nu \partial_\rho \psi+\partial_z\nu \partial_z\psi+(1/4)\exp(-2\nu+2\psi)[ (\partial_\rho \omega)^2+(\partial_z \omega)^2]~~~\cr
&+\Lambda\big[2\exp(2\mu+2\nu)/D_1+ 2 \exp(2\mu+2\psi)\omega^2(\partial_\rho \psi \partial_z \omega-\partial_\rho \omega \partial_z \psi)/(D_1\, D_2)\big]~~~,\cr
\end{align}

\begin{align}\label{psi}
\nabla^2\psi=&-\partial_\rho \psi (\partial_\rho \nu + \partial_\rho \psi)-\partial_z \psi (\partial_z\nu + \partial_z \psi)
-(1/2)\exp(-2\nu+2\psi)[(\partial_\rho \omega)^2+(\partial_z \omega)^2]~~~\cr
&+\Lambda\big[\exp(2\mu+2\nu)/D_1 + \exp(2\mu+2\psi) \omega^2 [-\partial_z \omega ( \partial_\rho \nu -2 \partial_\rho \psi) +
\partial_\rho \omega (\partial_z \nu - 2 \partial_z \psi)]/(D_1\, D_2)\big]~~~,\cr
\end{align}

\begin{align}\label{nu}
\nabla^2 \nu=&-\partial_\rho \nu (\partial_\rho \nu + \partial_\rho \psi) - \partial_z \nu (\partial_z\nu + \partial_z \psi)
+(1/2) \exp(-2\nu+2\psi)[(\partial_\rho \omega)^2+(\partial_z \omega)^2] \cr
&+\Lambda\big[-3\exp(2\mu + 2\nu)/D_1 + \exp(2\mu + 2 \psi) \omega^2 [-\partial_z \omega (\partial_\rho \nu + 2 \partial_\rho  \psi) + \partial_\rho \omega (\partial_z \nu + 2 \partial_z \psi)]/(D_1 \, D_2)\big]~~~.\cr
\end{align}

In writing these equations we have abbreviated
\begin{align}\label{d1d2}
D_1=&[\exp(2\nu)-\exp(2\psi)\omega^2]^3~~~,\cr
D_2=&\partial_\rho \omega \partial_z \nu-\partial_\rho\nu \partial_z \omega~~~.\cr
\end{align}
As checks on the manipulations leading to Eqs. \eqref{Omeg}--\eqref{nu}, (i) as noted above,  we checked that
the quasi-isotropic Kerr metric line element given in Appendix A.1 obeys the empty space equations obtained by dropping the dark energy terms with coefficient $\Lambda$ , and (ii) we checked that the spherically symmetric specialization of the axial formulas given in Appendix A.2 reduces to equations that both numerically and algebraically reproduce the results for the Schwarzschild-like case given in (AR).

Since we expect the solution to have reflection symmetry $z \leftrightarrow -z$, it suffices to solve these equations in the domain $z\geq 0$, imposing the condition that $\partial_z \omega=\partial_z  \mu = \partial_z \psi = \partial_z \nu =0$ for all $\rho$ on the axis where $z=0$.
For boundary conditions at ``infinity'', we require that sufficiently far from the origin, say $\rho \geq \rho_0$ and $z\geq z_0$ for some
specified $\rho_0$ and $z_0$, the solutions should agree with the leading large distance terms in the Kerr metric when written in quasi-isotropic
coordinates.
Thus at large distances we require
\begin{align}\label{larger}
\omega \simeq & 2M a/r^3 ~~~,\cr
e^{2\mu} \simeq & 1 + 2 M/r~~~,\cr
e^{2\psi} \simeq & \rho^2 (1+2M/r)~~~,\cr
e^{2\nu} \simeq & 1-2M/r ~~~,\cr
r=&(\rho^2 + z^2)^{1/2}~~~,\cr
\end{align}
with $M$ the ``black'' hole mass and $a$ the ``black'' hole angular momentum per unit mass.  At $\rho_0$, Eq. \eqref{larger} and its first partial derivative with respect to $\rho$ should be used as boundary conditions, in lieu of a boundary condition on the rotation axis $\rho=0$.\footnote{In (AR) we stated that in the spherical case we had to use the leading large $r$ term plus first order corrections in $\Lambda$ as the large distance boundary condition. On recalculating now,  we find that substantially identical results are obtained for the parameter values used in (AR) when the first order corrections in $\Lambda$ are dropped in the large distance boundary condition.}

\subsection{Residual general coordinate invariance of quasi-isotropic coordinates}

 The line element of Eq. \eqref{statmetric} has a residual general coordinate invariance consisting of a conformal mapping in the $\rho, z$ plane,
 \begin{equation}\label{conformal}
 \rho=f(\rho',z')~, ~~~ z=g(\rho',z').
 \end{equation}
 A standard result \cite{chandra} is that this preserves the form of Eq. \eqref{statmetric}  when
 \begin{equation}\label{invariance}
 f_{,1} = g_{,2}~~~~~~,~~~~~ f_{,2}=-g_{,1}~~~,
 \end{equation}
 which implies that both $f$ and $g$ are harmonic, $\nabla'^2f=\nabla'^2g=0$, and gives for the mapping
 \begin{align}\label{trans}
 e^{2\mu} (d\rho^2 + dz^2) \to & e^{2\mu'}(d\rho'^ 2+ dz'^2)~~~~,\cr
  e^{2\mu'}=& e^{2\mu}X~,\cr
  X=& f_{,1}^2+f_{,2}^2 = g_{,1}^2+g_{,2}^2= f_{,1} g_{,2}-f_{,2}g_{,1}~~~.\cr
 \end{align}
 The final line of Eq. \eqref{trans} shows that the positive quantity $X$ is also the Jacobian of the transformation.

 For the partial derivatives with respect to $\rho'$ and $z'$ we find from Eq. \eqref{conformal} that,
 \begin{align}\label{derividen}
 \frac{\partial}{\partial \rho'}=&\frac{\partial \rho}{\partial \rho'}\frac{\partial}{\partial \rho}
 +\frac{\partial z}{\partial \rho'}\frac{\partial}{\partial z}= f_{,1} \frac{\partial}{\partial \rho}
 + g_{,1}\frac{\partial}{\partial z}~~~,\cr
 \frac{\partial}{\partial z'}=&\frac{\partial \rho}{\partial z'}\frac{\partial}{\partial \rho}
 +\frac{\partial z}{\partial z'}\frac{\partial}{\partial z}= f_{,2} \frac{\partial}{\partial \rho}
 + g_{,2}\frac{\partial}{\partial z}~~~.\cr
 \end{align}
 Using these, and Eq. \eqref{invariance}, a short calculation gives
\begin{equation}\label{laplacian1}
\nabla'^2 = X \nabla^2~~~,
\end{equation}
\begin{equation}\label{dotprod}
\frac{\partial A}{\partial \rho'}\frac{\partial B}{\partial \rho'}+
\frac{\partial A}{\partial z'}\frac{\partial B}{\partial z'}=
X\left( \frac{\partial A}{\partial \rho}\frac{\partial B}{\partial \rho}+
\frac{\partial A}{\partial z}\frac{\partial B}{\partial z}\right)~~~,
\end{equation}
and
\begin{equation}\label{crossprod}
\frac{\partial A}{\partial \rho'}\frac{\partial B}{\partial z'}-
\frac{\partial A}{\partial z'}\frac{\partial B}{\partial \rho'}=
X\left( \frac{\partial A}{\partial \rho}\frac{\partial B}{\partial z}-
\frac{\partial A}{\partial z}\frac{\partial B}{\partial \rho}\right)~~~.
\end{equation}
In Eqs. \eqref{dotprod} and \eqref{crossprod}, $A$ and $B$  denote arbitrary functions of $\rho'$ and $z'$.

Using Eqs. \eqref{trans}--\eqref{crossprod}, one easily verifies
that the  differential equations Eqs. \eqref{Omeg}--\eqref{d1d2} are invariant in form
under the coordinate transformation of Eq. \eqref{trans}, while the boundary conditions of Eq. \eqref{larger} are not
invariant and serve to specify the coordinate choice uniquely. This makes numerical solution of the differential equations tricky.
For example, the Mathematica solver NDSolve returns a warning that the equations are ``convective'' and likely to be unstable, which we suspect may be a reflection of the invariance of Eq. \eqref{trans} that is only broken by the boundary conditions.  We leave the issue of how to solve the axial case equations for further study; it appears  that it may require development of a special purpose program, as opposed to using a commercial differential equation solver such as that in Mathematica.

\subsection{A mathematical question: When can a positive planar function be represented as the gradient squared of a harmonic function, or equivalently, as the modulus squared of a holomorphic function?}

The invariance demonstrated in Section IIIB poses the following question, which we discussed at length with several mathematician colleagues:
Given a two dimensional positive-definite\footnote{The positive semi-definite case is ruled out, as noted by  Helmut Hofer \cite{hofer} and Camillo De Lellis \cite{lellis}, because the continuation properties of harmonic functions imply that if X vanishes on an open set, it vanishes everywhere.} function $X(\rho,z)> 0$, can one find a harmonic function
$\nabla^2 f(\rho,z)=(\partial_\rho^2+\partial_z^2)f(\rho,z)=0$ such that $X$ is its squared gradient, $X(\rho,z)=(\partial_\rho f)^2 + (\partial_z f)^2$~?

A necessary condition noted by Camillo  De Lellis \cite{lellis} is that since f is harmonic, one has
\begin{equation} \label{ineq}
\nabla^2 X= \partial_i \partial_i (\partial_j f)^2 = 2 (\partial_i \partial_j f)^2 \geq 0,
\end{equation}
that is, $X$ must be subharmonic.  A stronger condition pointed out by  Terence Tao \cite{tao} is that ``Letting $g$ be the harmonic conjugate of $f$, we have from the Cauchy-Riemann equations that $\log X = 2 \log |f'+ig'|$ must be harmonic (except at zeroes), so this is a necessary condition, possibly also sufficient.''  Since
\begin{equation}\label{tao}
0=\nabla^2 \log X = (\nabla^2 X)/X - (\partial_i X)^2/X^2,
\end{equation}
the condition $\nabla^2 \log X=0$ implies the inequality of Eq. \eqref{ineq}.

To give a formal argument, we
 begin by noting following De Lellis \cite{lellis} that the conditions $X(\rho,z)=(\partial_i f)^2$ with $f$ harmonic, and $X(\rho,z)=|h|^2$ with $h$ a holomorphic function of $\rho+iz$, are equivalent.  For $f$ harmonic,  the complex function defined by $h=h_R+ih_I\equiv \partial_\rho f - i\partial_z f$ obeys the Cauchy-Riemann equations $\partial_\rho h_R=\partial_z h_I~,~\partial_z h_R=-\partial_\rho h_I$, and so is holomorphic with respect to the complex variable $\rho+iz$, and has the magnitude $X^{1/2}$.  Conversely, if $h$ is holomorphic the Cauchy-Riemann equations are  integrability conditions which imply that
$h_R$ and $h_I$ are gradients of a harmonic``potential'' $f(\rho,z)$, that is $h_R=\partial_\rho f~,~~h_I=-\partial_z f$.

Following Tao \cite{tao}, taking the logarithm of $h$ we have
\begin{equation}\label{log1}
\log h=\frac{1}{2}\log X+i \arg\,h~~~,
\end{equation}
and since the logarithm of a holomorphic function is also holomorphic, we have new Cauchy-Riemann equations
\begin{align}\label{log2}
\partial_\rho \frac{1}{2} \log X= &\partial_z \arg\,h~~~,\cr
\partial_z \frac{1}{2} \log X=&-\partial_\rho \arg\,h~~~.\cr
\end{align}
Differentiating the first line with respect to $\rho$, the second line with respect to $z$,  adding and doubling we get
\begin{equation}\label{adding}
\partial_\rho^2 \log X +\partial_z^2 \log X=\nabla^2 \log X=0~~~,
\end{equation}
which is the asserted necessary condition.  Following De Lellis \cite{lellis}, this argument can be run in reverse.  In the language of
differential forms, Eq. \eqref{adding} states that the differential form
\begin{equation}\label{form}
\chi:=\partial_\rho \log X dz- \partial_z \log X d\rho
\end{equation}
is closed,
\begin{equation}\label{closed}
d\chi=\partial_\rho^2 \log X d\rho dz-\partial_z^2 \log X dz d\rho = \nabla^2 \log X d\rho dz=0~~~.
\end{equation}
On a simply connected domain, this implies by the Poincar\'e lemma that $\chi$ is exact, which implies Eq. \eqref{log2}, and therefore the
existence of a holomorphic function $h$ of which $X$ is the absolute value squared.  This shows that on a simply connected domain the
condition of Eq. \eqref{adding} is also a sufficient condition.

As noted by Tao \cite{tao}, When the domain is not simply connected, the variation (holonomy) of $\arg h$ around a closed curve $\gamma$ not passing through singularities
is a multiple of $2 \pi i$, but not necessarily zero.  From Eq. \eqref{log2}, we have
\begin{equation}\label{lineint}
\int_\gamma \frac{1}{2} (\partial_\rho \log X dz -\partial_z \log X d\rho)= 2 \pi n~~~
\end{equation}
for some integer $n$,
or equivalently,
\begin{equation}\label{lineint1}
\int_\gamma \frac{\partial_\rho X dz-\partial_z X d\rho}{X}  =4\pi n~~~.
\end{equation}
Assuming there are no essential singularities, this condition gives an additional constraint on $X$, which asserts that the line integral of Eq. \eqref{lineint1} is quantized, and  the order of blowup of $X$ around each singularity has to be an even integer.

\subsection{A `` no-go'' result for the attempted reduction of the four-function axial line element to three-function form}

We now use the mathematical results of the preceding section to show that the invariance  exhibited in Section IIIB cannot be eliminated by using it to reduce the four function axial line element to a three function form.

 We begin by remarking that
the inverse transformation to Eq. \eqref{conformal} can be written as
\begin{equation}\label{inverse}
\rho'=F(\rho,z)  ~~~~~,~~~~~z'=G(\rho,z)~~~,
\end{equation}
and since the inverse of a conformal map is also conformal, $F$ and $G$ are both harmonic, a fact that will be used later on.
Applying the transformation of Eq. \eqref{trans} to the line element of Eq. \eqref{statmetric},  the effect of the mapping is to replace
$2\mu$ in final term by $2\mu + \log X$, and since the theorem of Section IIIC implies that  $\log X$ is harmonic, the covariance group of the line element consists
of adding a harmonic function to $\mu$. Thus the solution to the differential equation system is made unique by  boundary conditions on $\mu$ sufficient to uniquely specify a harmonic function.

 What would be nice is to fix $X$ by imposing the coordinate condition
\begin{equation}
e^{2\mu} X = e^{2 \psi} /\rho'^{\,2} \equiv e^{2\kappa},
\end{equation}
so that the line element becomes
\begin{equation}\label{new}
ds^2= (- e^{2\nu}+\omega^2 \rho'^{\,2} e^{2\kappa}) dt^2 - 2 e^{2\kappa} \rho'^{\,2} \omega dt d\phi
+ e^{2\kappa} (d\rho'^{\,2} + dz'^{\,2}+\rho'^{\,2} d\phi^2)~~~,
\end{equation}
which apart from the rotation $\omega$ has a conformal spatially Euclidean form.  This involves only three unknown functions $\nu, \kappa, \omega$ instead of four, and if attainable could simplify numerical calculations using the line element of Eq. \eqref{statmetric}.  Unfortunately, we shall see that in general this simplification is not possible, as exemplified by the specific example of the Kerr metric.

To make this transformation, we need to impose  \big(with $\rho'=F(\rho,z)$ \big)
\begin{equation}
 \log X= 2 (\psi-\mu)-2\log  F,
\end{equation}
which using the condition $\nabla^2 \log X=0$ gives
\begin{align}\label{fequation}
\nabla^2 \log F=& \nabla^2(\psi-\mu)\cr
 =&-Y~~~,\cr
Y=& 2 [\nu_{,1}(\rho, z) \psi_{,1}(\rho, z)+\nu_{,2}(\rho, z) \psi_{,2}(\rho, z)]\cr
  + &\psi_{,1}(\rho, z)^2 + \psi_{,2}(\rho, z)^2+
(3/4)  e^{-2 \nu(\rho, z) +
   2 \psi(\rho, z)} [\omega_{,1}(\rho, z)^2 + \omega_{,2}(\rho, z)^2]
    ~~~,\cr
\end{align}
where we have substituted the empty space form of the Einstein equations for $\psi$ and $\mu$; when matter or dark energy is present there will be additional terms in $Y$.

Since $\nabla^2 F=0$,
\begin{equation}\label{fcond}
Y = - \nabla^2 \log F = (\partial_i F)^2/F^2 \geq 0,
\end{equation}
that is, $Y$ must be positive semi-definite.   An additional condition can be found by further application of the result of Sec. IIIC.    Rewriting Eq. \eqref{fcond} as
\begin{equation}\label{logcond}
\log Y= \log(\partial_i F)^2-2\log F~~~,
\end{equation}
and noting that since $F$ is harmonic, the results of Section IIIC imply that $\nabla^2  \log(\partial_i F)^2 =0$.
Therefore Eq. \eqref{logcond} implies
\begin{equation}\label{nablacond}
\nabla^2 \log Y= -2 \nabla^2 \log F ~~~,
\end{equation}
and using  Eq. \eqref{fequation} this becomes a condition on $Y$,
\begin{equation}\label{ycond}
\nabla^2 \log Y=2 Y~~~.
\end{equation}
The problem formulated in Section IIIB is to analyze a deformation of the Kerr metric resulting from the dark energy action of Eq. \eqref{eff}, with a boundary condition that the modified metric should  asymptote to the Kerr metric.  A numerical study of the Kerr metric using Mathematica shows that
$Y$ for Kerr is indeed everywhere non-negative, but a numerical check on the condition of Eq. \eqref{ycond} at a few sample points shows that it
is {\it not} satisfied.  Therefore for both Kerr and the deformation under study,  a conformal transformation cannot be used to reduce the original four-function line element of Eq. \eqref{statmetric} to the three-function form of Eq. \eqref{new}, which if possible would have
simplified the problem of finding a stable numerical solution.

\section{Open issues}

The exposition of the preceding sections leaves a number of evident open issues.  Here are a few, more are given in the review \cite{adreview}  :
\begin{itemize}

\item  What is the physics governing the ``black'' hole interior in the Schwarzschild-like case? Can one prove that the modified ``black'' hole leaks particles?   How does one calculate the leakage rate? We plan to first address these questions in the computationally tractable spherical case.
\item  Can a phenomenological parametrization of the leakage be used to explain some of the puzzling features of the vicinity of the ``black'' hole at the center of our Galaxy?  See \cite{ak} for preliminary ideas.  Similarly, can it be used to help model galaxy formation, where preliminary ideas are given in \cite{gal}.
\item What is the best way to numerically solve the axial case equations?  Would minimization of a cost function formed from the differential equations and boundary conditions work?
\item  Does the solution in the axial case suggest a possible role in  jet formation  or other directional phenomena?

Further study of these issues is planned.

\end{itemize}

\section{Acknowledgements}
I wish to thank Helmut Hofer, Camillo De Lellis, and Terence Tao for incisive private communications on the mathematical question discussed in Section IIIC, as specifically referenced there. I wish to thank Fethi Ramazano\u glu for a helpful email on boundary conditions, and the referee for helpful suggestions on presentation.   The work of this paper was performed in part with benefit from the hospitality of
the Aspen Center for Physics, which is supported by the National Science Foundation grant PHY-1607611.

\appendix

\section{Formlas used for checking the axial-vector equations}

\subsection{The Kerr metric in quasi-isotropic coordinates}

We give here the Kerr metric after transformation  \cite{brandt}, \cite{allen}  from Boyer-Lindquist to quasi-isotropic coordinates.  In terms of the cylindrical coordinates $\rho$,  $z$, and  the spherical coordinate radius  $r=(\rho^2+z^2)^{1/2}$,
auxiliary quantities $\chi^2$, $\tilde\rho^2$, $\Delta$, and $\Sigma$ are given by
\begin{align}\label{auxquant}
\chi^2=&[1+(M+a)/(2r)][1+(M-a)/(2r)]~~~,\cr
\tilde\rho^2=&r^2 \chi^4 +a^2 z^2/r^2~~~,\cr
\Delta=& r^2 \chi^4-2 M r \chi^2  + a^2~~~,\cr
\Sigma=& (r^2 \chi^4+a^2)^2-\Delta \, a^2 \rho^2/r^2~~~.\cr
\end{align}
In terms of these, the functions appearing in the Kerr line element when put in the quasi-isotropic form of Eq. \eqref{statmetric} are given by
\begin{align}\label{Kerrfun}
e^{2\nu}=&1-2 M r \chi^2/\tilde\rho^2 +(2 M \rho \chi^2 a)^2 /(\tilde\rho^2 \Sigma)~~~,\cr
e^{2\psi}=& \Sigma \rho^2/(\tilde\rho^2  r^2)~~~,\cr
e ^{2\mu}=&\tilde\rho^2/r^2~~~,\cr
\omega=&2 M r \chi^2 a/\Sigma~~~.\cr
\end{align}
We have checked these formulas by using Mathematica to show that they obey Eqs. \eqref{Omeg}--\eqref{nu} when the
dark energy terms with coefficient $\Lambda$ and denominators $D_1$ and $D_2$  are dropped.
Expanding Eqs. \eqref{Kerrfun} to leading order in $M$ gives the formulas used as boundary conditions in Eq. \eqref{larger}.

\subsection{Spherically symmetric specialization of the axial equations}

The spherical reduction of Eq. \eqref{statmetric} is obtained by specializing to
\begin{equation}\label{spherred}
\omega=0~,~~\mu=\mu(r)~,~~\nu=\nu(r)~,~~\psi=\mu(r)+\log \rho~~~.
\end{equation}
Using $\partial_\rho r=\rho/r$ and $\partial_z r=z/r$, we find that the axial differential equations reduce to (with $'$ denoting $d/dr$)
\begin{align}\label{spherred1}
\mu''=&\nu'\mu'+(1/r)(\nu'-\mu')+2e^{2\mu-4\nu}~~~,\cr
\mu''=&-(3/r)\mu'-(1/r)\nu'-\mu'(\mu'+\nu')+e^{2\mu-4\nu}~~~,\cr
\nu''=&-(2/r)\nu'-\nu'(\nu'+\mu')-3e^{2\mu-4\nu}~~~.\cr
\end{align}
Averaging the two equations for $\mu''$ gives the alternative equation
\begin{equation}\label{simpler}
\mu''=-(1/2)(\mu')^2-(2/r)\mu'+1.5e^{2\mu-4\nu}~~~.
\end{equation}
Numerically solving Eq. \eqref{simpler} for $\mu''$ together with  Eq. \eqref{spherred1}  for $\nu''$ gives results for the spherical isotropic coordinate case in agreement with those of \cite{adler2}, and we also checked that these formulas reduce algebraically to those used in \cite{adler2}.


\begin{thebibliography}{99}
\bibitem{chandra}  S. Chandrasekhar, ``The Mathematical Theory of Black Holes'', Clarendon Press, Oxford, 1983 (1992, 2000), especially Chapters 2 and 6.
\bibitem{pani} V. Cardoso and P. Pani, Living Rev. Relativ. {\bf 22}, 4 (2019).
\bibitem{adler1} S. L. Adler, Classical Quantum Gravity {\bf 30}, 195015 (2013), arXiv:1306.0482.
\bibitem{adler2} S. L. Adler and F. M. Ramazano\u glu, Int. J. Mod. Phys. D {\bf 24}, 1550011 (2015), arXiv:1308.1448.
\bibitem{adler3} S. L. Adler, Int. J. Mod. Phys. D {\bf 25} 1643001 (2016), arXiv:1605.05217.
\bibitem{adler4} S. L. Adler, Int. J. Mod. Phys. D {\bf 26},1750159 (2017), arXiv:1704.00388.
\bibitem{adler5} S. L. Adler, Phys. Rev. D {\bf 100}, 123503 (2019), arXiv:1905.08228.
\bibitem{adler6} S. L. Adler, Int. J. Mod. Phys. D {\bf 30}, 2150044 (2021), arXiv:2008.07598.
\bibitem{thooft}  G. 't Hooft, Int. J. Mod. Phys. D {\bf 24}, 1543001 (2015), arXiv:hep-th/1410.6675.
\bibitem{adreview}  S. L. Adler, Modern Physics Letters A {\bf 36}, 2130027 (2021), arXiv:2111.12576.
\bibitem{hawking}  S.W. Hawking and G. F. R. Ellis, {\it The large scale structure of space-time}, Cambridge Monographs on Mathematical Physics, Cambridge University Press (1973), p. 320.
\bibitem{faraoni} V. Faraoni, Galaxies {\bf 1}, 114 (2013); arXiv:gr-qc/1309.4915.
\bibitem{krishnan1} B. Krishnan, ``Quasi-local black hole horizons'', in Springer Handbook of Spacetime, A. Ashtekar and V. Petkov, eds., Springer Verlag (2014), pp. 527-555; arXiv:gr-qc/1303.4635.
\bibitem{adler7}  S. L. Adler, Int. J. Mod. Phys. D {\bf 30}, 2142023 (2021), arXiv:2105.07521.
\bibitem{nielsen} A. B. Nielsen and M. Visser, Class. Quantum Grav. {\bf 23}, 4637 (2006); arXiv:gr-qc/0510083.
\bibitem{ashtekar} A. Ashtekar, C. Beetle, and J. Lewandowski,  Class. Quantum Grav. {\bf 19}, 1195 (2002); arXiv:gr-qc/0111067.
\bibitem{krishnan2} A. Ashtekar and B. Krishnan,  Living. Rev. Rel. {\bf 7}:10 (2004); arXiv:gr-qc/0407042.

\bibitem{lu}  J. R. Lu {\it et al.},  ApJ {\bf 625}, L51 (2005);  also ApJ {\bf 690}, 1463 (2009), and the online article at the following URL:   http://black holes.stardate.org/research/milky-way-star-clusters.php.html
\bibitem{lu1}  T. R. Lauer {\it et al.}, AJ  {\bf 116}, 2263 (1998); R. Bender {\it et al.}, ApJ {\bf 631}, 280 (2005); A.C. Seth {\it et al.}, AJ 132, 2539 (2006); J. R. Lu {\it et al.}, ApJ {\bf 764}, 155 (2013).
\bibitem{ak} S. L. Adler and K. Singh, ``A One-Dimensional Model for Star Formation Near a `Leaky' Black Hole'', arXiv:2112.12319.
\bibitem{gal} S. L. Adler, ``A mechanism for a `leaky' black hole to catalyze galaxy formation'', arXiv:2112.12491.
\bibitem{bland} R. D. Blandford and R. L. Znajek, Mon. Not. Roy. Astron. Soc.  {\bf 179}, 433 (1977).
\bibitem{brandt} S. R. Brandt and E. Seidel, Phys. Rev. D {\bf 52}, 856 (1995), arXiv:gr-qc/9412072;  Phys. Rev. D {\bf 54}, 1403 (1996), arXiv:gr-qc/9601010.
\bibitem{allen} J. Allen, ``Analytic Kerr Solution for Puncture Evolution'', available at URL:  https://wwwrel.ph.utexas.edu/Members/jon/papers/punctureKerr.pdf
\bibitem{hofer}  H. Hofer, private email communication to the author.
\bibitem{lellis} C. De Lellis, private email communications to the author.
\bibitem{tao} T. Tao, private email communications to the author.

\end{thebibliography}
\end{document}